\documentclass[a4paper]{article}

\pdfoutput=1

\usepackage[utf8]{inputenc}
\usepackage[T1]{fontenc}
\usepackage[english]{babel}
\usepackage{csquotes}

\usepackage{amsmath, amsfonts, bm}

\usepackage[a4paper,top=3cm,bottom=2cm,left=3cm,right=3cm,marginparwidth=1.75cm]{geometry}
\usepackage{parskip}
\usepackage{booktabs}
\usepackage{tabularx}
\usepackage{multirow}
\usepackage{makecell}
\usepackage{graphicx}
\usepackage{subfig}

\usepackage[version=4]{mhchem}
\usepackage{siunitx}
\DeclareSIUnit\angstrom{\text{Å}}

\usepackage{xcolor}
\usepackage{textcomp}
\usepackage[normalem]{ulem}
\usepackage{outlines}

\usepackage{authblk}

\usepackage[backend=biber, natbib=true, style=numeric, sorting=none]{biblatex}
\usepackage{nomencl}
\usepackage[final]{changes}
\usepackage{hyperref}
\hypersetup{
    colorlinks=true,
    allcolors=blue
}

\date{}

\title{Towards a fully differentiable digital twin for solar cells}

\author[1,2,$\dagger$]{Marie Louise Schubert}
\author[1,2,$\dagger$]{Houssam Metni}
\author[1]{Jan David Fischbach}
\author[1]{Benedikt Zerulla}
\author[14]{Marjan Krsti\'{c}}
\author[3,4]{Ulrich W. Paetzold}
\author[3,4]{Seyedamir Orooji}
\author[7]{Olivier J. J. Ronsin}
\author[7,8]{Yasin Ameslon}
\author[7,8,9]{Jens Harting}
\author[10,11]{Thomas Kirchartz}
\author[10]{Sandheep Ravishankar}
\author[10]{Chris Dreessen}
\author[10]{Eunchi Kim}
\author[4,5]{Christian Sprau}
\author[4,6]{Mohamed Hussein}
\author[4,5]{Alexander Colsmann}
\author[7]{Karen Forberich}
\author[12,13]{Klaus J\"ager}
\author[1,2,$\star$]{Pascal Friederich}
\author[1,14,$\star$]{Carsten Rockstuhl}

\affil[1]{Institute of Nanotechnology, Karlsruhe Institute of Technology (KIT), Karlsruhe, Germany}
\affil[2]{Institute of Theoretical Informatics, Karlsruhe Institute of Technology (KIT), Karlsruhe, Germany}
\affil[3]{Institute of Microstructure Technology, Karlsruhe Institute of Technology (KIT), Karlsruhe, Germany}
\affil[4]{Light Technology Institute, Karlsruhe Institute of Technology (KIT), Karlsruhe, Germany}
\affil[5]{Material Research Center for Energy Systems, Karlsruhe Institute of Technology (KIT), Karlsruhe, Germany}
\affil[6]{Department of Physics, Faculty of Science, Ain Shams University, Cairo, Egypt}
\affil[7]{Helmholtz-Institute Erlangen-Nürnberg for Renewable Energy (IET-2), Forschungszentrum Jülich, Erlangen, Germany}
\affil[8]{Department of Chemical and Biological Engineering, Friedrich-Alexander-Universität Erlangen-Nürnberg, Erlangen, Germany}
\affil[9]{Department of Physics, Friedrich-Alexander-Universität Erlangen-Nürnberg (FAU), Erlangen, Germany}
\affil[10]{IMD-3 Photovoltaik, Forschungszentrum Jülich (FZJ), Jülich, Germany}
\affil[11]{Faculty of Electrical Engineering and Information Technology, University of Duisburg-Essen, Duisburg, Germany}
\affil[12]{Department Optics for Solar Energy (SE-AOPT), Helmholtz-Zentrum Berlin für Materialien und Energie GmbH, Berlin, Germany}
\affil[13]{Zuse Institute Berlin, Berlin, Germany}
\affil[14]{Institute of Theoretical Solid State Physics, Karlsruhe Institute of Technology (KIT), Karlsruhe, Germany}

\affil[$\dagger$]{These authors contributed equally; marie.schubert@kit.edu and houssam.metni@kit.edu}
\affil[$\star$]{These authors are corresponding authors; pascal.friederich@kit.edu and carsten.rockstuhl@kit.edu}

\begin{document}
\maketitle
\clearpage
\begin{abstract}
Maximizing energy yield (EY) -- the total electric energy generated by a solar cell within a year at a specific location -- is crucial in photovoltaics (PV), especially for emerging technologies. Computational methods provide the necessary insights and guidance for future research. However, existing simulations typically focus on only isolated aspects of solar cells. This lack of consistency highlights the need for a framework unifying all computational levels, from material to cell properties, for accurate prediction and optimization of EY prediction. To address this challenge, a differentiable digital twin, $\text{Sol}(\text{Di})^2\text{T}$, is introduced to enable comprehensive end-to-end optimization of solar cells. The workflow starts with material properties and morphological processing parameters, followed by optical and electrical simulations. Finally, climatic conditions and geographic location are incorporated to predict the EY. Each step is either intrinsically differentiable or replaced with a machine-learned surrogate model, enabling not only accurate EY prediction but also gradient-based optimization with respect to input parameters. Consequently, $\text{Sol}(\text{Di})^2\text{T}$ extends EY predictions to previously unexplored conditions. Demonstrated for an organic solar cell, the proposed framework marks a significant step towards tailoring solar cells for specific applications while ensuring maximal performance. 
\end{abstract}

\section{Introduction}\label{section: intro}

Modern photovoltaic applications for building integration or in agricultural settings require solar cell technologies with customizable properties. Emerging PV technologies offer a promising solution to this need~\cite{emerging_device_performance}. Among the many research directions in the context of emerging PV technologies, organic (OPV) and perovskite solar cells stand out~\cite{Jaeger_overview_TWphotovoltaics}. These PV technologies feature mechanical flexibility, lightweight design and have the potential for cost-effective production~\cite{Overview_OPV, green2014emergence, OPV_printing, ColsmanetalOPV}. The cells can be made transparent and in tunable colors~\cite{OPV_color}, expanding their applicability in design-sensitive contexts. Furthermore, some of these technologies have demonstrated competitive power conversion efficiencies (PCE)~\cite{BallifHaug_NatureReview, HurniHaug_perovskite_tandem, WilsonCatchpole_2020}.

In recent years, the energy yield (EY) has become an important technological objective in the PV research community. The EY encompasses the total electric energy generated by solar cells within a year in a given location, accounting for external factors such as irradiance and temperature~\cite{Schmager:19}. Computational tools have been developed to better understand PV processes and to predict the EY for specific solar cell settings~\cite{schmager_2021_4696257, McCallum_AWalker_2024, PerformanceModelTopic2010, hussein2024device, Courtier_WalkerA_perovskitePCE}. Despite this progress, many existing computational tools are developed independently and only focus on isolated aspects of solar cells. Therefore, a comprehensive material-to-application EY simulation framework that targets specific materials, processes, and device characteristics is highly desirable~\cite{VOGT2022111944, TopicUlbrich2025}.

Digital twins offer an innovative and impactful approach to address this challenge~\cite{LIU2021346}.
We define a digital twin as a holistic simulation workflow that encompasses all relevant components and scales of a physical system, which are typically considered separately, and includes capabilities for design and performance enhancement~\cite{tao2018digital}. Our digital twin not only simulates the system’s behavior but also enables virtual optimization, allowing for iterative improvements and decision-making. 
While digital twins are widely adopted in industry, they have also been applied to various physics-based problems in academic research~\cite{GUNASEGARAM2021102089, YI2023107203, gu2025}. 
In photovoltaics, Ng Wei Tat {\it et al.} designed a digital twin to model the roll-to-roll OPV printing process, integrating surrogate machine learning (ML) models into the optimization loop to enhance the PCE of printed solar cells~\cite{NG2024102038}. More recently, Lüer {\it et al.} discussed the potential of digital twins to address key challenges in photovoltaics~\cite{lueer_digital_twin}. 

The purpose of our work is to introduce a digital twin framework that bridges the gap between currently isolated methods and establishes a holistic EY calculation framework. The digital twin supports data-driven decision-making and accelerates the development and optimization of next-generation PV technologies. In the future, experimental and simulation domains refine one another, enabling calibration of model parameters, enhancing predictive capability under previously unexplored operating conditions, and ultimately facilitating the realization of devices with maximized EY. 

Building on this, we present a \textbf{Sol}ar \textbf{Di}fferentiable \textbf{Di}gital \textbf{T}win, that we designate as \textbf{$\text{Sol}(\text{Di})^2\text{T}$}, combining physical simulations and ML to calculate and virtually optimize the EY based on material properties, process parameters, device architecture, and placement at specific locations.

$\text{Sol}(\text{Di})^2\text{T}$ aims for full differentiability, meaning the entire computational framework is structured to allow for the calculation of gradients, thereby transforming the digital twin from a mere predictive simulator into an inverse-design tool capable of performing gradient-based optimization to identify ideal parameters which maximize the EY.
While previous research took the first steps along such a path, \textit{e.g.}, Mann {\it et al.} developed a differentiable solar cell simulator focusing on drift-diffusion simulation of electrical properties~\cite{MANN2022108232}, 
our approach is comprehensive and incorporates more simulation aspects across the different length scales of a solar cell. These are discussed in the Methodology Section~\ref{section: methods}. We choose to demonstrate our digital twin for an organic solar cell. 
Although implemented for a single instance, $\text{Sol}(\text{Di})^2\text{T}$ is versatile and can accommodate other solar cell designs by adjusting the underlying methodology. This work provides a solid foundation for future research, providing a comprehensive understanding of the EY across different solar cell parameters. $\text{Sol}(\text{Di})^2\text{T}$ is open-source and implemented in Python, with the full code accessible on GitHub (\url{https://github.com/aimat-lab/SolDi2T}). Our efforts contribute to a long-term endeavor to accelerate knowledge transfer between scientific research and industrial production, ultimately bridging the gap between first-principle calculations and experimental validation.

\section{Methodology}\label{section: methods}

\subsection{Setting}\label{subsection:setting}
In $\text{Sol}(\text{Di})^2\text{T}$, we develop and use simulation tools and ML methods to model solar cell properties at different levels. Moreover, we implement our framework in a differentiable manner. \textbf{Figure~\ref{fig:figure1}}(a) shows the five-step scheme of our digital twin. We start with the morphological description of the material, which is then passed on, along with the layer stack, to the optical simulations, where charge generation rates are calculated. Next, the electrical simulations are performed. In a final step of forward passing, the EY of the given solar cell is determined. With the help of ML-based optimization tools, we can close the loop and optimize the solar cell parameters. The differentiable formulation allows us to compute the derivatives of our objective function, \textit{i.e.}, the EY, with respect to the degrees of freedom that characterize the solar cell. These derivatives facilitate gradient-based local optimization, enabling the iterative refinement of parameters to approach the optimal configuration.

In the present work, we apply $\text{Sol}(\text{Di})^2\text{T}$ to an organic solar cell with \textbf{PM6:Y6} (details of the molecules can be found in Section S~1 of the Supporting Information). PM6:Y6 is an excellent organic photoactive material~\cite{YUAN20191140}. The overall stack architecture of the example is shown in {Figure~\ref{fig:figure1}(b). When light hits the sample stack, it enters through the glass substrate. We assume the glass substrate to be \qty{1}{\micro\meter}-thick. The substrate carries a transparent indium tin oxide electrode (\textbf{ITO}, \qty{150}{\nano\meter}) and an electron transport layer of zinc oxide (\textbf{ZnO}, \qty{30}{\nano\meter}). 
The light then passes through the photoactive material. Initially, we choose the photoactive material PM6:Y6 to be \qty{200}{\nano\meter}-thick. The semi-transparent back contact of the solar cell comprises silver nanowires (\textbf{AgNW}) dispersed in a layer of poly(3,4-ethylenedioxythiophene):poly(styrene sulfonate) (\textbf{PEDOT:PSS}, \qty{190}{\nano\meter}). For simplicity reasons, the AgNWs are assumed to be a homogenized slab with a \qty{10}{\nano\meter}-thin layer of Ag in the layer of PEDOT:PSS.

\begin{figure} [ht]
  \centering
  \begin{minipage}[t]{.9\linewidth}
      {\includegraphics[width=1.0\linewidth]{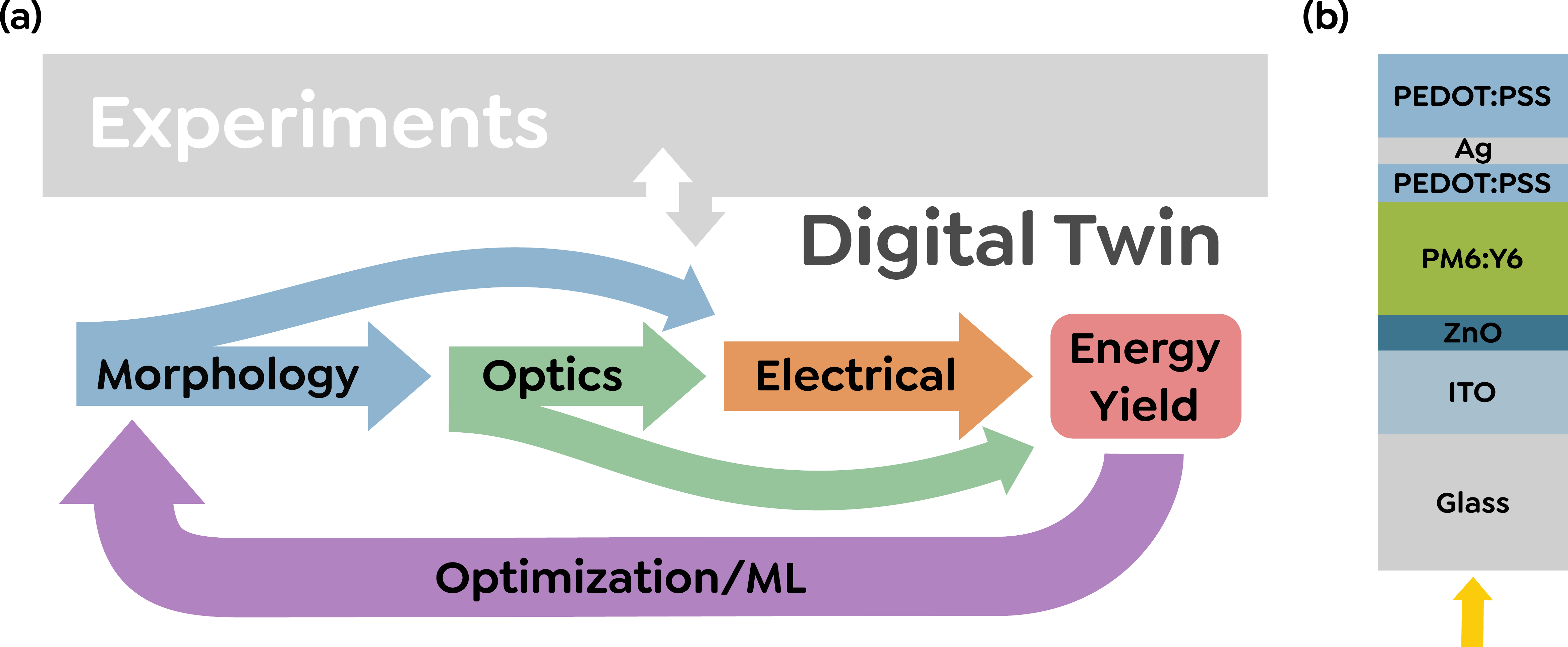}}%
  \end{minipage} 
  \hfill
  
  \caption
    {\textbf{(a)} Overview of the digital twin framework $\text{Sol}(\text{Di})^2\text{T}$ for energy yield (EY) optimization, comprising five simulation stages: material morphologies, optical properties, electrical properties, EY calculation framework, and optimization and machine learning (ML). \textbf{(b)} Device architecture of the studied cell. The different thicknesses are not drawn to scale. The exact stack thicknesses are \qty{150}{\nano\meter} of ITO, \qty{30}{\nano\meter} of ZnO, \qty{200}{\nano\meter} of PM6:Y6, \qty{50}{\nano\meter} of PEDOT:PSS, a layer of \qty{10}{\nano\meter} of Ag, and finally a layer of \qty{140}{\nano\meter} of PEDOT:PSS. The yellow arrow represents the incoming light.}
  \label{fig:figure1}
\end{figure}

The following subsections outline the methods and computational tools employed at various stages of the workflow, each addressing a distinct aspect of solar cell properties. These aspects include the simulation of materials' morphology, optical, and electrical properties, an EY calculation framework, and a final ML and optimization stage. The details of each forward-passed computational method can be found in \textbf{Figure~\ref{fig:digital_twin_detailed}}.

\begin{figure} [ht]
  \centering
  \begin{minipage}[t]{0.99\linewidth}
      {\includegraphics[width=1.0\linewidth]{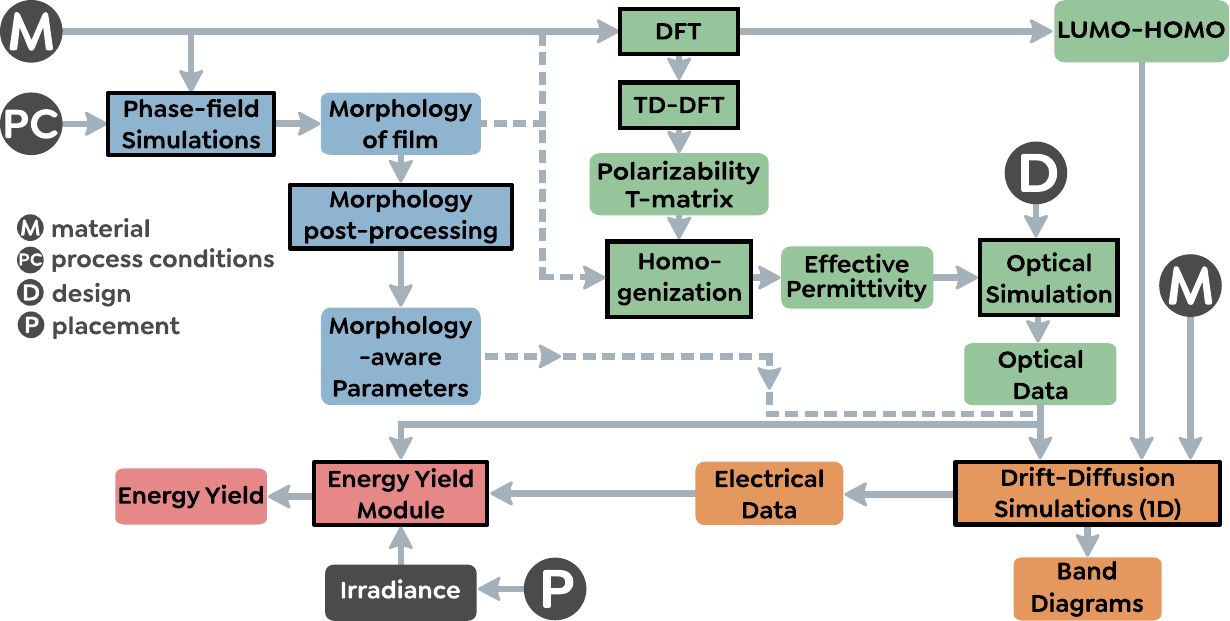}}
  \end{minipage} 
  \hfill
  \caption
    {Detailed framework of the forward-pass of the digital twin $\text{Sol}(\text{Di})^2\text{T}$. The different sub-structures are given the same colors as in Figure~\ref{fig:figure1}: material morphologies (blue), optical properties (green), electrical properties (orange), and the energy yield calculation framework (red). The input quantities (in circles) are material (M), process conditions (PC), design (D), and placement (P). Material (M) encompasses material properties and experimental knowledge of the material. Process conditions (PC) include temperature and drying conditions. Design (D) is defined by the details of the layer stack of the solar cell. Placement (P) includes the location and orientation of the solar cell.  
    Squared boxes, framed in black, are methods, and rounded, unframed patches are (intermediate) results.
    }
  \label{fig:digital_twin_detailed}
\end{figure}%

\subsection{Morphology}\label{subsection: morphology}
The starting point of the $\text{Sol}(\text{Di})^2\text{T}$ workflow is the simulation of the morphology formation of the photoactive layer of the solar cell. Information about the bulk-heterojunction (BHJ) nanostructure is crucial because its properties strongly impact the final optoelectronic properties of the PV device. As illustrated in Figure~\ref{fig:digital_twin_detailed} in blue, the inputs required by the morphology part of our digital twin are material properties (thermodynamic and kinetic parameters), and the film processing conditions (solvents, temperature, drying conditions). The simulations allow us to describe how crystallization and/or demixing are triggered upon solvent evaporation, develop during drying, and are ultimately quenched when the film is dry. The output is a full spatially resolved description of the BHJ composition and crystallinity. This information can subsequently be used in the optical and electrical modules of the digital twin to obtain morphology-aware light absorption properties, IV curves, and ultimately the EY.

The simulations are performed using a phase field (PF) model developed to investigate the thin film structuring in printed PV~\cite{ronsin_phase-field_2020,ronsin_phase-field_2021,ronsin_phase-field_2022,ronsin_role_2020,steinberger_challenges_2024}. The PF model is a continuum mechanics approach that simulates the kinetic evolution of a possibly multiphase and multimaterial system towards its thermodynamic equilibrium~\cite{warren_phase_2009,michels_predictive_2021,kim_modeling_2009,takaki_phase-field_2014,ronsin_phase-field_2022}. PF simulations handle the length scales (from nanometers to microns) and time scales (up to seconds or even minutes) that are relevant for the BHJ formation of solution-processed solar cells with tractable computational effort~\cite{warren_phase_2009,michels_predictive_2021,kim_modeling_2009,takaki_phase-field_2014}.
Thereby, the thermodynamic properties of the studied PV system are encoded in a Gibbs free energy functional. Our method handles physical processes involved in the BHJ formation, in particular solvent evaporation, liquid demixing, and crystallization~\cite{ronsin_phase-field_2020,ronsin_phase-field_2021,ronsin_phase-field_2022,ronsin_role_2020,steinberger_challenges_2024}. The forces driving the system towards its thermodynamic equilibrium are derived from the Gibbs free energy functional. The kinetics of phase changes (both evaporation and crystallization) are described by Allen-Cahn equations coupled to Cahn-Hilliard equations describing diffusive mass transport. During all the simulation steps, we ensured that the PF model is coupled to a fluid dynamics solver, which takes advective mass transport into account. See the Supporting Information Section~S2 for details of the simulation method.

\subsection{Optics}\label{subsection: optics}

Once the film morphology is known, the next step is the optical analysis of the solar cell. Such analysis requires computing the electromagnetic response of the solar cell to determine the fraction of absorbed photons in the photoactive layer. As illustrated in green in Figure~\ref{fig:digital_twin_detailed}, the optical calculations take two major inputs: the molecules making up the photoactive material and the overall solar cell architecture.
The optical properties (the complex refractive index or electric permittivity) of the molecular components are obtained using quantum chemical methods: ``Density-functional theory'' (DFT) for ground state properties and linear response, ``Time-dependent density-functional theory'' (TD-DFT) for excited state properties such as molecular polarizabilities. The polarizabilities then enable the construction of the transition matrix (T-matrix)~\cite{Waterman_Tmatrix}, describing light-matter interactions at the molecular level~\cite{Benedikt_multiscale, Benedikt_homogenization}. Next, we translate the properties of the individual molecules to a dielectric function that can be used to study the light-matter interaction at the continuum level. In this so-called homogenization step, the effective T-matrix (\textrm{\textbf{T}$_{\textrm{eff}}$}) for the bulk material is computed using the \emph{treams} framework~\cite{treams_Beutel}. \textrm{\textbf{T}$_{\textrm{eff}}$} enables determining key material parameters of the bulk photoactive material, such as permittivity \textrm{$\bm{\varepsilon}_{\textrm{eff}}$} and permeability \textrm{$\bm{\mu}_{\textrm{eff}}$}. Unlike other homogenization approaches, this method does not need any prior knowledge about the shape of the bulk material~\cite{Benedikt_homogenization}.

Using the calculated material properties, optical simulations analyze the light propagation in solar cells. Depending on the complexity of the considered solar cells, either numerical optical simulations (\textit{e.g.}, COMSOL Multiphysics, JCMsuite) or analytical methods, such as the transfer matrix method~\cite{TMM_OHeavens_1960}, can be used. We model the solar cell as a stack of stratified isotropic media and use a semi-analytical stack solver based on scattering matrices to study the optical response~\cite{Sax_Github}. In this approach, optical properties, such as the total absorption, are calculated by successively combining layer scattering matrices. This means that photon absorption is quantified within each solar cell layer and across the entire stack. This enables analysis and optimization of material properties, thicknesses, and layer arrangements. Further details of this simulation step are provided in the Supporting Information Section~S3. 
The current framework assumes an internal quantum efficiency (IQE) of 100~\%, that is, all photons absorbed in the photoactive layer are converted into free charge carriers. This assumption holds if domain sizes remain small, and complete exciton dissociation occurs. Section~\ref{subsection: morphology} will indeed show that all phases in the considered cell exhibit small domain sizes. The final outcome of the optical simulation step is the charge generation rate that is fed to the electrical simulation part of our digital twin. 

To enable differentiable modeling of the optical response, a surrogate ML model is employed to approximate the charge generation rate obtained from scattering matrix simulations. Specifically, a neural network is trained to predict the generation rate as a function of incident angle, photoactive layer thickness, and wavelength. Further details on the neural network surrogate model are provided in the Supporting Information Section~S3.

\subsection{Electrical} 
\label{subsection:electrical}
With the generation rate as input from the optical calculations, electrical simulations are performed as the next element in the framework of the digital twin, as illustrated in orange in Figure~\ref{fig:digital_twin_detailed}. The aim of this module is to provide current-density (\textit{JV}) data as a function of the solar cell temperature, irradiance, and photoactive layer thickness, which are forwarded to the EY calculations in the next step. The electrical simulations of the organic solar cell were carried out using an effective medium drift-diffusion model~\cite{Koster2005PRB, kirchartz2014device}. 
In our approach, the bulk heterojunction is modeled in the same way as an inorganic semiconductor, where the lowest unoccupied molecular orbital (LUMO) of the acceptor replaces by the conduction band and the highest occupied molecular orbital (HOMO) of the donor replaces by the valence band. The energy difference between the LUMO of the acceptor and the HOMO of the donor thereby replaces the traditional role of the band gap. Due to the disordered nature of molecular semiconductors, the precise band gap energy is more challenging to define than in crystalline semiconductors~\cite{Koopmans22solrrl}, as the localized states near the band edges often require modeling with exponential or Gaussian band tails~\cite{Kirchartz11prb}. However, the development of modern donor and acceptor molecules for organic photovoltaics has led to sharper band tails that reduce the uncertainty in the band gap energies~\cite{Kaiser2021}. In inorganic semiconductors, non-radiative band to band recombination is negligible due to the negligible reorganization energy. Thus, in most inorganic thin-film semiconductors, Shockley-Read-Hall (SRH) recombination via localized states is dominant~\cite{shockley1952statistics}. This is different in molecular semiconductors, which feature a finite reorganization energy in the excited charge-transfer state at the donor-acceptor interface, leading to non-zero rates of non-radiative bimolecular recombination even in the absence of additional defect states~\cite{benduhn2017intrinsic,azzouzi2018nonradiative}. Thus, it is possible to model recombination using an effective bimolecular recombination coefficient that primarily includes non-radiative contributions as well as an additional SRH term for deeper localized states at the donor-acceptor interface. This second term is important in some situations and allows the ideality factor to exceed unity~\cite{kirchartz2013differences,saladina2023power, sandberg2023mid,zeiske2021direct}. 

The drift-diffusion simulations are performed using the software Advanced Semiconductor Analysis (ASA)~\cite{pieters2006advanced}, which numerically solves three coupled differential equations in one dimension, namely the Poisson equation and the continuity equations for electrons and holes (refer to the Supporting Information Section~S4). To achieve realistic input variables for the model, experimental temperature- and irradiance-dependent \textit{JV} parameters were fitted with a common parameter set~\cite{saladina2024transport}.

To create a differentiable version of the drift-diffusion simulator, we use simulated \textit{JV} curves as training data for a neural network. A \textit{JV} curve dataset is generated over a range of input parameters using ASA. These include the temperature $T$ in $\mathrm{K}$, the irradiance in $\mathrm{suns}$, the layer thickness in $\mathrm{nm}$, the injection barriers $\phi_\mathrm{bf}$ and $\phi_\mathrm{bb}$ at the front and back contact as well as the bandgap energy $E_\mathrm{g}$ in $\mathrm{eV}$, the zero-field mobility $\mu_{0}$ in $\mathrm{cm}^{2}\mathrm{V}^{-1}\mathrm{s}^{-1}$ and the radiative recombination coefficient $k_\mathrm{rec,0}$ in $\mathrm{cm}^{3}\mathrm{s}^{-1}$. 
A multi-output neural network is then trained on this dataset.  
For the subsequent calculation of the EY in the workflow, only a temperature-, irradiance-, and layer thickness-dependent open-circuit voltage $V_\mathrm{oc}$ and fill factor $FF$ are required. 
Therefore, the remaining parameters are fixed.
Further details are presented in the Supporting Information Section~S4. 

\subsection{Energy Yield}

The last part of the forward model of $\text{Sol}(\text{Di})^2\text{T}$, shown in red in Figure~\ref{fig:digital_twin_detailed}, is the calculation of the EY. In this work, we use EYCalc~\cite{schmager_2021_4696257}, an open-source EY modeling software designed for detailed solar cell analysis. It takes the optical data and the electrical data as input parameters. A comprehensive description is provided by~\textcite{Schmager:19}.
EYCalc, with capabilities validated in literature~\cite{Chung2017, Gota2020, Lehr2020, DeBastiani2021, Gota2022, Hu2024}, also enables the simulation of EY for a wide range of PV system architectures under realistic irradiation conditions. These include complex-textured and planar designs, monofacial and bifacial configurations, as well as single- and multi-junction PV devices.

The original EYCalc software consists of four core modules: the irradiance module, the optics module, the electrical module, and the EY module. Recently, a degradation sub-module was incorporated into the electrical module to account for the impact of degradation on the long-term EY of solar cells~\cite{Orooji2024}.
In this work, we modify the software to better suit our needs (see Supporting Information Section~S5). The irradiance module retains its original purpose, computing direct and diffuse irradiance spectra for each hour of the year using typical meteorological year (TMY3) data for various locations across the USA~\cite{Wilcox2008}. These spectra are generated by feeding meteorological data into the Simplified Model of Atmospheric Radiative Transfer of Sunshine (SMARTS)~\cite{GUEYMARD2001325}, followed by a simple cloud model to account for weather effects. By integrating the optical and electrical data (described in the previous Sections), we calculate the EY of the solar cell, taking into account its orientation (rotation and/or tilt) and specific location.
While the original implementation of the EYCalc software was in MATLAB, it is adapted to Python using the JAX framework to make it differentiable for $\text{Sol}(\text{Di})^2\text{T}$~\cite{jax2018github}. JAX enables the efficient, automatic computation of high-order gradients for complex numerical functions, and is a natural choice in this case. 

\subsection{Optimization}

By combining the previously presented simulation methods, $\text{Sol}(\text{Di})^2\text{T}$ provides a unifying framework for a holistic calculation of the EY. For a clear overview of the input parameters used for the digital twin, refer to Table~S5 in the Supporting Information. 

Once the EY can be accurately evaluated, it can be optimized with respect to architecture design and material parameters.
Several optimization strategies have been used in literature for various PV-related tasks, such as design of experiments combined with machine learning~\cite{Cao2018, Kirkey2020}, Bayesian optimization~\cite{Zhan2024, McCallum2023}, and gradient-based approaches~\cite{Ismaeel2021, Premkumar2023, Hassan2021}.
Among these, gradient-based approaches are particularly efficient, especially in computational settings, offering fast convergence, fewer evaluations, and strong suitability for high-dimensional, smooth optimization problems such as EY maximization.

Building on these advantages, our approach $\text{Sol}(\text{Di})^2\text{T}$ is designed to be differentiable, facilitating gradient-based optimization through backpropagation.
The gradient information propagated through each interconnected block allows to efficiently optimize the EY. 
Critical factors such as material composition, process conditions, and device configurations can be optimized by calculating the gradients of EY with respect to various input parameters.
While some simulations in the earlier steps are inherently differentiable due to the algorithms and methods employed, others are not easily differentiable. We propose leveraging ML models as surrogates to address this challenge and replace these non-differentiable simulations. This involves generating a dataset of simulation points spanning a broad range of the input parameter space and training a supervised ML model. The resulting surrogate model seamlessly integrates into the differentiable workflow, enabling efficient gradient computation. This strategy is used when simulating optical and electrical properties of the solar cell, as detailed in Subsections~\ref{subsection: optics} and~\ref{subsection:electrical}.

\section{Results and Discussion}\label{section: discussion}  

In the following, we present the results of each forward step of our digital twin $\text{Sol}(\text{Di})^2\text{T}$. We apply it to the organic solar cell that is described in Section \ref{subsection:setting}. As a result, device-specific aspects become increasingly relevant. Furthermore, we demonstrate the optimization process and its results. 

\subsection{Morphology}
\label{subsection: morphology_results}

In $\text{Sol}(\text{Di})^2\text{T}$, we perform phase field simulations of the PM6:Y6 photoactive layer with the mass blend ratio 1:1.2 and with processing conditions that favour the formation of PM6 and Y6 crystals~\cite{EES_reference_2021, AM_reference_massratio} (simulation parameters given in the Supporting Information Section~S2). The donor and acceptor volume fraction and crystalline order parameter results obtained from the PF simulations are shown in \textbf{Figure~\ref{fig:PhaseType}}. The volume fraction is a conservative physical parameter that corresponds locally to the ratio between a component’s volume inside a voxel and a voxel volume. The crystalline order parameter is a non-conservative physical parameter, describing the crystallinity of a species, ranging from 0 for a fully amorphous component to 1 for a fully crystalline component.

Data from the crystal formation of PM6 and Y6 is used to identify the various coexisting phases based on their composition and crystallinity (See Figure~\ref{fig:PhaseType}). In the present example, four phases are identified: donor crystals (high donor crystalline order and volume fraction), acceptor crystals (high acceptor crystalline order and volume fraction), a mixed phase (low crystalline order for both donor and acceptor, intermediate volume fractions), and an amorphous donor phase (high volume fraction and low crystalline order of donor). All phases exhibit small domain sizes below \qty{40}{\nano\meter}, consistent with experimental findings~\cite{yuan_single-junction_2019,yang_intrinsic_2023,liu_tuning_2021}. Their respective fractions and average compositions are shown in Table~\ref{tab:phasetype}. Given that domain sizes are sufficiently small relative to the optical wavelength, we can treat the material as isotropic and homogeneous in the subsequent optical analysis to capture its effects while enabling a simplified description.

In the near future, the analysis of the simulated BHJ morphology will be extended to calculate morphology-aware descriptors of exciton dissociation, of non-geminate recombination, and of charge mobilities in the photoactive layer (see Supporting Information Section~S2). These descriptors will then be used to calculate the corresponding inputs of the electrical model.

\begin{figure}[h!]
     \centering
    \includegraphics[width=\textwidth]{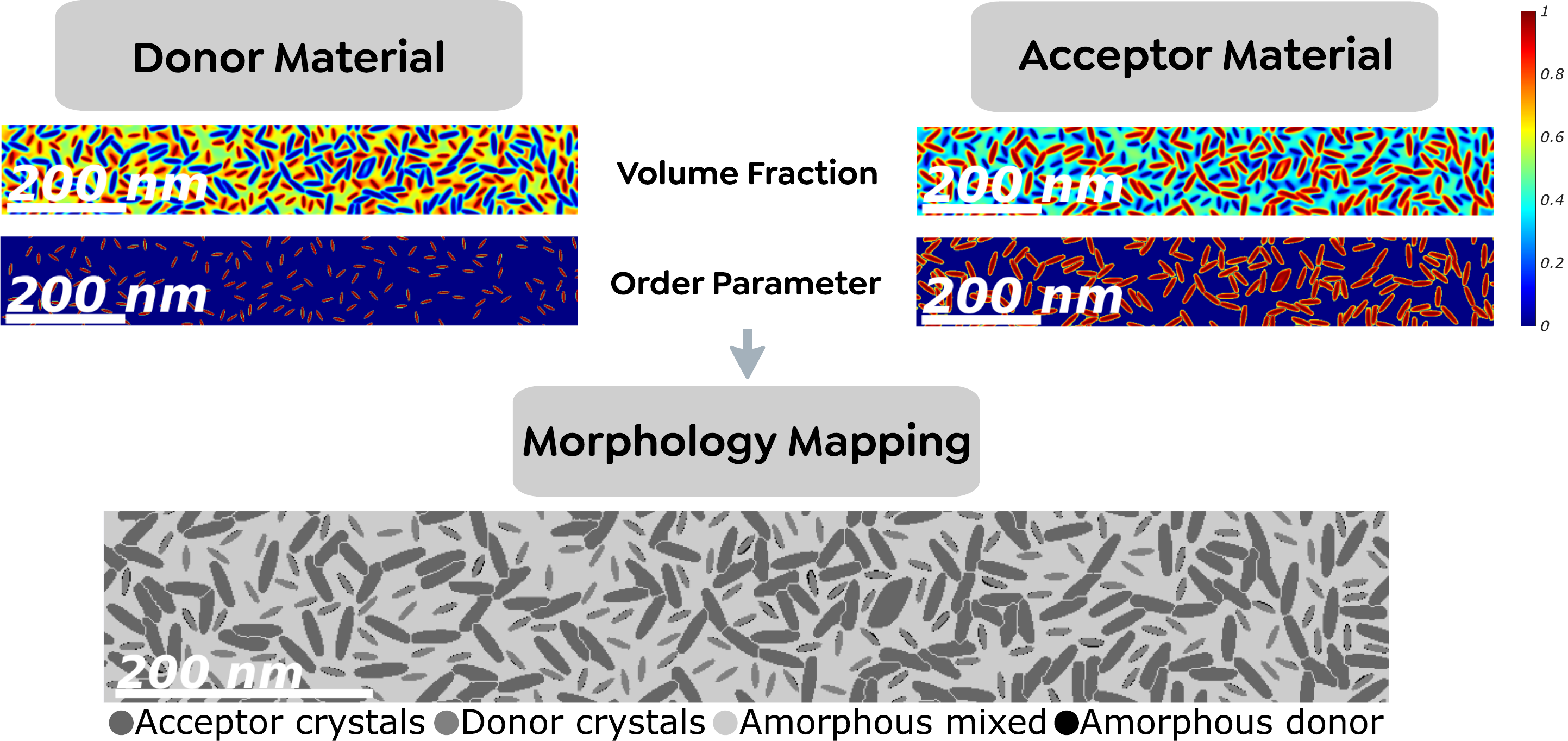}  
\caption{Morphology module results. The top part shows the BHJ morphology of PM6:Y6 from PF simulations. The volume fraction and crystalline order parameter are depicted for both the donor and acceptor material. The volume fraction and crystalline order parameter spread from 0 or low values (dark blue) to 1 or higher values (dark red). 
The bottom part shows the morphology mapping we attain from the donor (PM6) and acceptor (Y6) material. It illustrates the four coexisting phases in the BHJ, which are listed in Table~\ref{tab:phasetype}. The phases are: donor crystal phase (dark grey), acceptor crystal phase (medium grey), amorphous mixed phase (light grey), and amorphous donor phase (black).}  
     \label{fig:PhaseType}
 \end{figure}

\begin{table}[h!]
\centering
\resizebox{\textwidth}{!}{
\begin{tabular}{lcccc}
\toprule
\textbf{Phase Property} 
& \textbf{Donor crystals} 
& \textbf{Acceptor crystals} 
& \textbf{Amorphous mixed} 
& \textbf{Amorphous donor} \\
\midrule
Proportion [\%]          
& 8.7 & 33.9 & 56.8 & 0.6 \\
Donor Volume Fraction 
& 0.97 & 0.16 & 0.63 & 0.92 \\
\bottomrule
\end{tabular}
}
\caption{Composition and proportion of the coexisting phases extracted from the simulated morphologies.}
\label{tab:phasetype}
\end{table}

\subsection{Optics}
The results of the optical simulations are provided in \textbf{Figure~\ref{fig:Rockstuhl_results}}.
Starting from the chosen finite-size molecular structure, DFT is used to determine the LUMO and HOMO gap, and as a basis for the TD-DFT calculations. These then allow us to calculate the dynamic polarizabilities and the T-matrix of the molecules. The outputs enable us to attain the optical properties via the T-matrix formalism.
The optical properties are homogenized to derive the optical properties of the bulk photoactive PM6:Y6 blend. The morphology analysis in Section~S\ref{subsection: morphology_results} indicates that the material can be treated as isotropic at the relevant optical length scales. Therefore, using a semi-analytical stack solver to study the optical response is sufficient. Combined with other material parameters defined by the device architecture, the charge generation rate of the whole stack is calculated (further details in Section~S3 of the Supporting Information). 

The charge generation rate of the photoactive material is higher compared to the other material layers of the stack and decreases with further depth. The rate is evaluated over a wavelength range from \qty{300}{\nano\meter} to \qty{1200}{\nano\meter} and can be computed for various light polarizations, angles of incidence, and material thicknesses. Figure~\ref{fig:Rockstuhl_results}(d) shows the charge generation rate within a stack of a \qty{200}{\nano\meter} thick photoactive layer at normal incidence, where p- and s-polarizations exhibit identical behavior. 
Training of the surrogate charge generation rate model achieves an $R^2$ score above 0.99 on test data, ensuring high fidelity to the original optical calculations. 
This charge generation rate is passed on to the subsequent calculations of the drift-diffusion and the EY.

\begin{figure} [ht]
  \centering
  \begin{minipage}[t]{0.85\linewidth}
      {\includegraphics[width=1.0\linewidth]{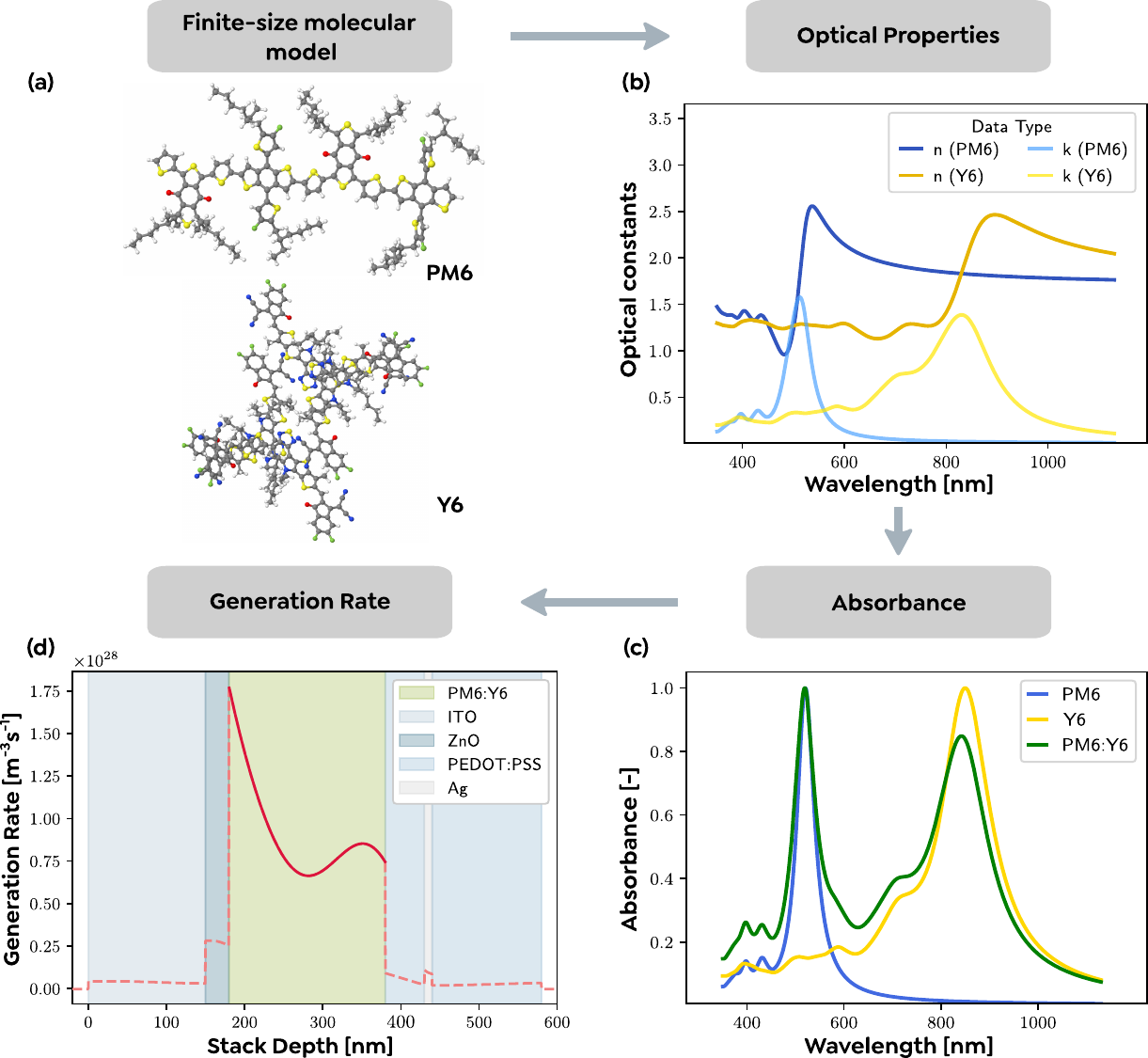}}%
  \end{minipage} 
  \hfill
  \caption
    {Optics module results. \textbf{(a)} The considered finite-sized molecular model of PM6:Y6. \textbf{(b)} The optical properties of the two molecules obtained from TD-DFT calculations. \textbf{(c)} The absorbance of the material mixture PM6:Y6 calculated via homogenization of the molecular optical properties. \textbf{(d)} The charge generation rate, where different colors indicate the different layers of the stack. The lime green color indicates the photoactive material PM6:Y6. 
    }
  \label{fig:Rockstuhl_results}
\end{figure}%

\subsection{Electrical}

Electrical simulations typically rely on a large number of input parameters. Low effective mobilities of electrons and holes, along with the high transition rates for non-radiative recombination, imply that charge transport and recombination parameters are the most important factors in determining the device performance in organic solar cells. In this regard, it is known that the charge carrier mobilities and non-radiative transition rates are strong functions of the temperature~\cite{tanase2003unification,benduhn2017intrinsic}. We therefore model the temperature dependence of the charge carrier mobility $\mu$ (electron and hole mobilities assumed to be equal) and the bimolecular recombination coefficient \textit{k}$_{rec}$ using Arrhenius equations, given by
\begin{equation}
k_{\mathrm{rec}}=k_{\mathrm{rec}, 0} \exp \left(\frac{-E_{\mathrm{A}, \mathrm{rec}}}{k_{\mathrm{B}} T}\right),
\end{equation}
\begin{equation}
\mu=\mu_0 \exp \left(\frac{-E_{\mathrm{A}, \mathrm{mob}}}{k_{\mathrm{B}} T}\right),
\end{equation}
where $E_\mathrm{A,rec}$ and $E_\mathrm{A,mob}$ are the respective activation energies, and $k_\mathrm{rec,0}$ and $\mu_0$ are the respective pre-factors.

To determine these activation energies and pre-factors, we use the temperature dependence of the open-circuit voltage ($V_\mathrm{oc,exp}$) and fill factor ($FF_\mathrm{exp}$) of optimized PM6:Y6 devices from the experimental current-voltage curves reported in~\cite{saladina2024transport}. \textbf{Figure~\ref{fig:SR01}}(a,b) illustrates the temperature-dependent $V_\mathrm{oc,exp}$ and $FF_\mathrm{exp}$ values for different light intensities. $V_\mathrm{oc,exp}$ and $FF_\mathrm{exp}$ are fitted using corresponding data obtained from drift-diffusion simulations of the current-voltage curves of a device with the structure cathode/PM6:Y6/anode (simulation parameters shown in Table~S2 in the Supporting Information). These fits yield values of $k_\mathrm{rec,0}=4.0\times10^{-11}$ cm$^3$s$^{-1}$ and $E_\mathrm{A,rec}=-56$ meV for the bimolecular recombination coefficient, and values of $\mu_0=3.7\times10^{-2}$ cm$^{2}$V${^-1}$s$^{-1}$ and $E_\mathrm{A,mob}=70.4$ meV for the mobility. The temperature-dependent evolution of the bimolecular recombination coefficient and the mobility is shown in Figure~\ref{fig:SR01}(c,d), which are used to generate the training data at different temperatures and for different PM6:Y6 thicknesses.
Training of the surrogate \textit{JV} curve neural network achieved high accuracies on unseen test data with $R^2$ scores of $0.96$ and $0.98$, respectively for the $V_\mathrm{oc}$ and $FF$.
The results are further developed in the Supporting Information Section~S4. 

\begin{figure} [h!]
    \centering
    \includegraphics[width=\textwidth]{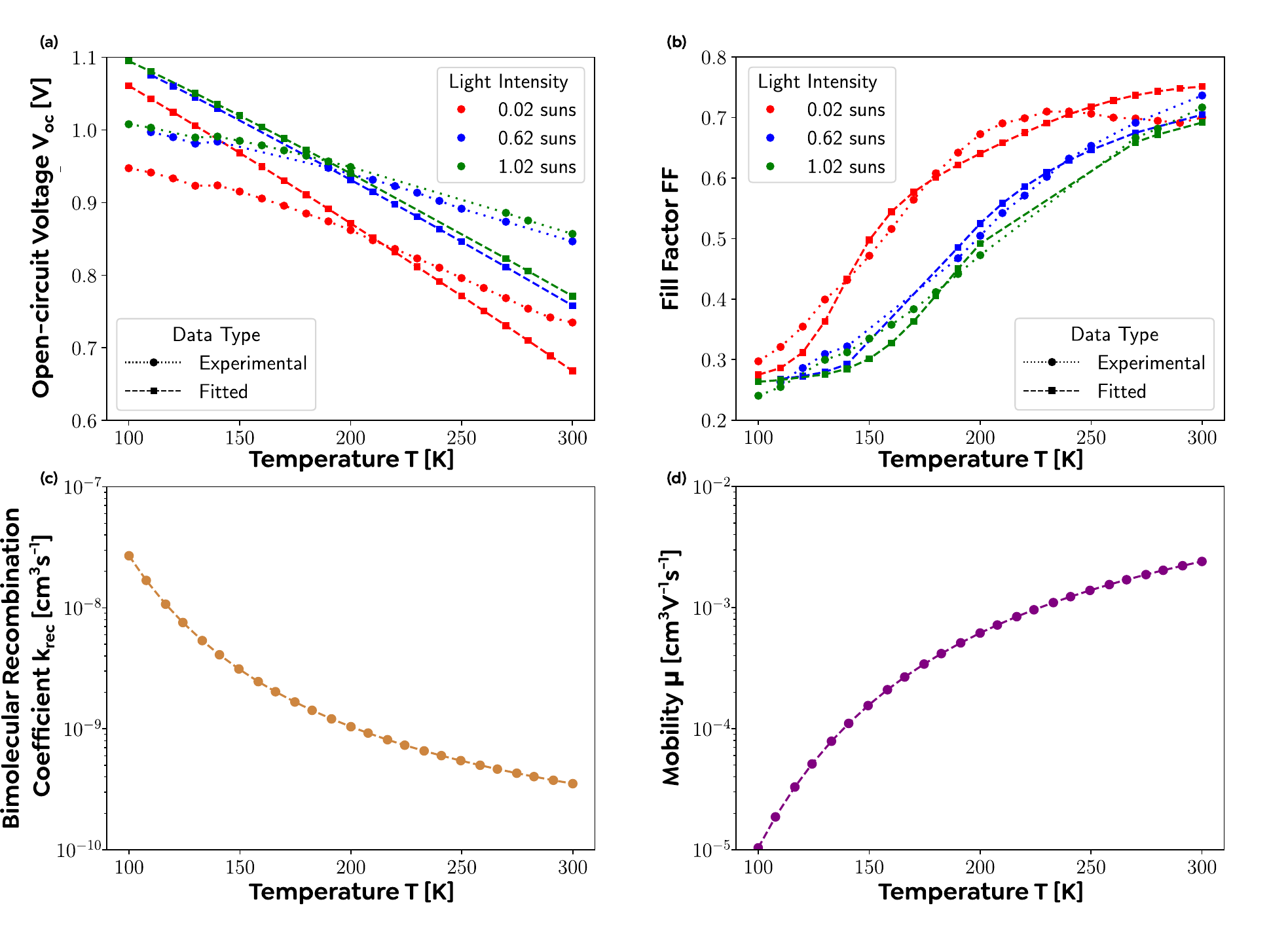}
    \caption{Electrical module results. \textbf{(a)} Open-circuit voltages for different light intensities as a function of temperature. The spheres represent experimental data and the squares the corresponding fits. \textbf{(b)} Fill factors for different light intensities as a function of temperature. The spheres represent experimental data and the squares the corresponding fits. \textbf{(c)} The obtained bimolecular recombination coefficient $k_\mathrm{rec}$ as a function of temperature. \textbf{(d)} The obtained mobility $\mu$ as a function of temperature. }
    \label{fig:SR01}
\end{figure}

\subsection{Energy Yield}
\label{subsection:ey_results}
In this final step of the forward-modeling part of $\text{Sol}(\text{Di})^2\text{T}$, we compare the EY of the PM6:Y6 solar cell for three different photoactive layer thicknesses (\qty{100}{\nano\meter}, \qty{200}{\nano\meter}, and \qty{300}{\nano\meter}) across three locations in the USA: Phoenix (Arid/Desert), Seattle (Temperate), and Honolulu (Tropical/Humid). As shown in \textbf{Figure~\ref{fig:EYResults}}, the solar cell with a photoactive layer thickness of \qty{200}{\nano\meter} consistently achieves the highest EY across all locations and tilt angles, while the \qty{300}{\nano\meter} photoactive layer thickness yields the lowest EY. The \qty{100}{\nano\meter} thickness also delivers a high EY, though slightly lower than the \qty{200}{\nano\meter} thickness. 

Using EYCalc, we show that the variation in the EY among the various thicknesses is attributed to the underlying solar cell parameters ($J_\mathrm{sc}$, $V_\mathrm{oc}$, and $FF$), evaluated under standard test conditions (STC). Although the \qty{300}{\nano\meter} device shows the highest $J_\mathrm{sc}$, its significantly reduced $FF$ leads to the lowest overall performance. In contrast, the \qty{200}{\nano\meter} device provides the best balance between $J_\mathrm{sc}$ and $FF$, resulting in the highest PCE, and therefore, the highest EY. The \qty{100}{\nano\meter} device performs intermediately, with the lowest $J_\mathrm{sc}$ due to lower absorption, but the highest $FF$. Detailed parameter values are provided in Table~S3 of the Supporting Information.

Additionally, the results highlight that Phoenix exhibits the highest EY overall, regardless of photoactive layer thickness or tilt angle, while Seattle has the lowest EY. This is due to the significantly higher solar irradiance in Phoenix compared to the other locations.

\begin{figure} [ht]
  \centering
  \begin{minipage}[t]{1.0\linewidth}
      {\includegraphics[width=\linewidth]{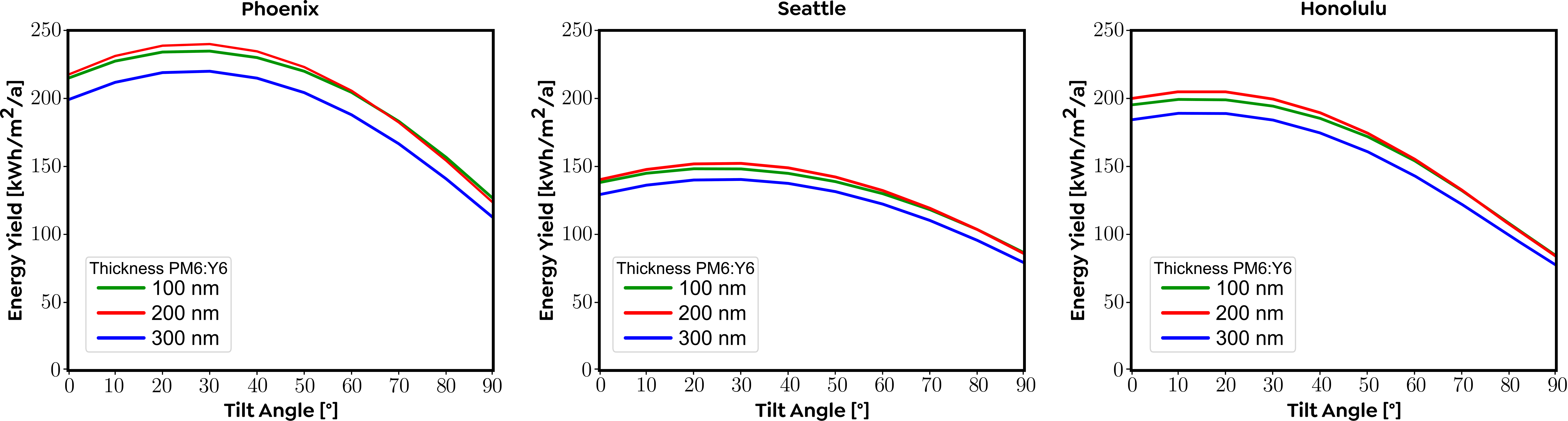}}%
  \end{minipage} 
  \hfill
  \caption
    {Energy yield (EY) module results. The different plots show the simulated EY values of the PM6:Y6 solar cell for varying photoactive layer thicknesses and tilt angles across different climate zones classified by the Köppen classification: Phoenix (Arid), Seattle (Temperate), and Honolulu (Tropical).}
  \label{fig:EYResults}
\end{figure}%

\subsection{Optimization}
To demonstrate the feasibility of inverse design using $\text{Sol}(\text{Di})^2\text{T}$ in a proof-of-principle test, we restrict the degrees of freedom to two parameters.
This simplification enables a systematic exploration of the parameter space while analyzing the EY, enabling us to visualize the full landscape of the objective function and assess the performance of the gradient-based inverse design method. The two chosen parameters for the optimization process are the photoactive layer thickness in $\mathrm{nm}$, included in the optical, electrical, and energy yield simulation blocks, and the tilt angle of the solar cell in degrees, involved in the EY calculation. 
The optimization is performed across the three locations presented in Subsection~\ref{subsection:ey_results}. The three factors, photoactive layer thickness, tilt angle, and location, have been shown to heavily influence solar cell and module performance~\cite{LIPOVSEK2022111421, Mehleri2010, vanBavel2009, Minnam2010, moule2006}.
The optimization is constrained within the bounds of [\qty{100}{\nano\meter}, \qty{300}{\nano\meter}]  and [\qty{0}{\degree}, \qty{90}{\degree}] for the photoactive layer thickness and the tilt angle, respectively. 

\textbf{Figure~\ref{fig:optimization}} presents gradient descent optimization results for EY in Phoenix. The figure presents a contour plot of the EY along with optimization trajectories originating from four distinct initial conditions—comprising all combinations of two thicknesses (\qty{125}{\nano\meter} and \qty{275}{\nano\meter}) and two tilt angles (\qty{10}{\degree} and \qty{50}{\degree}). The optimization trajectories show how the parameters are updated step by step during the optimization process. To provide further insight, arrows indicating the direction and magnitude of the objective function’s gradient are added. The arrows help visualize how the gradient descent algorithm navigates the parameter space, revealing why the optimization paths curve or converge in certain directions.

In addition to classical gradient descent, we apply a range of both gradient-based and gradient-free optimization methods to benchmark performance. These comparisons, detailed in Section~S6 of the Supporting Information, highlight the advantages of using a differentiable simulation framework: gradient-based methods consistently reach convergence in fewer function evaluations compared to their gradient-free counterparts.

\begin{figure} [ht]
  \centering
      {\includegraphics[width=0.8\linewidth]{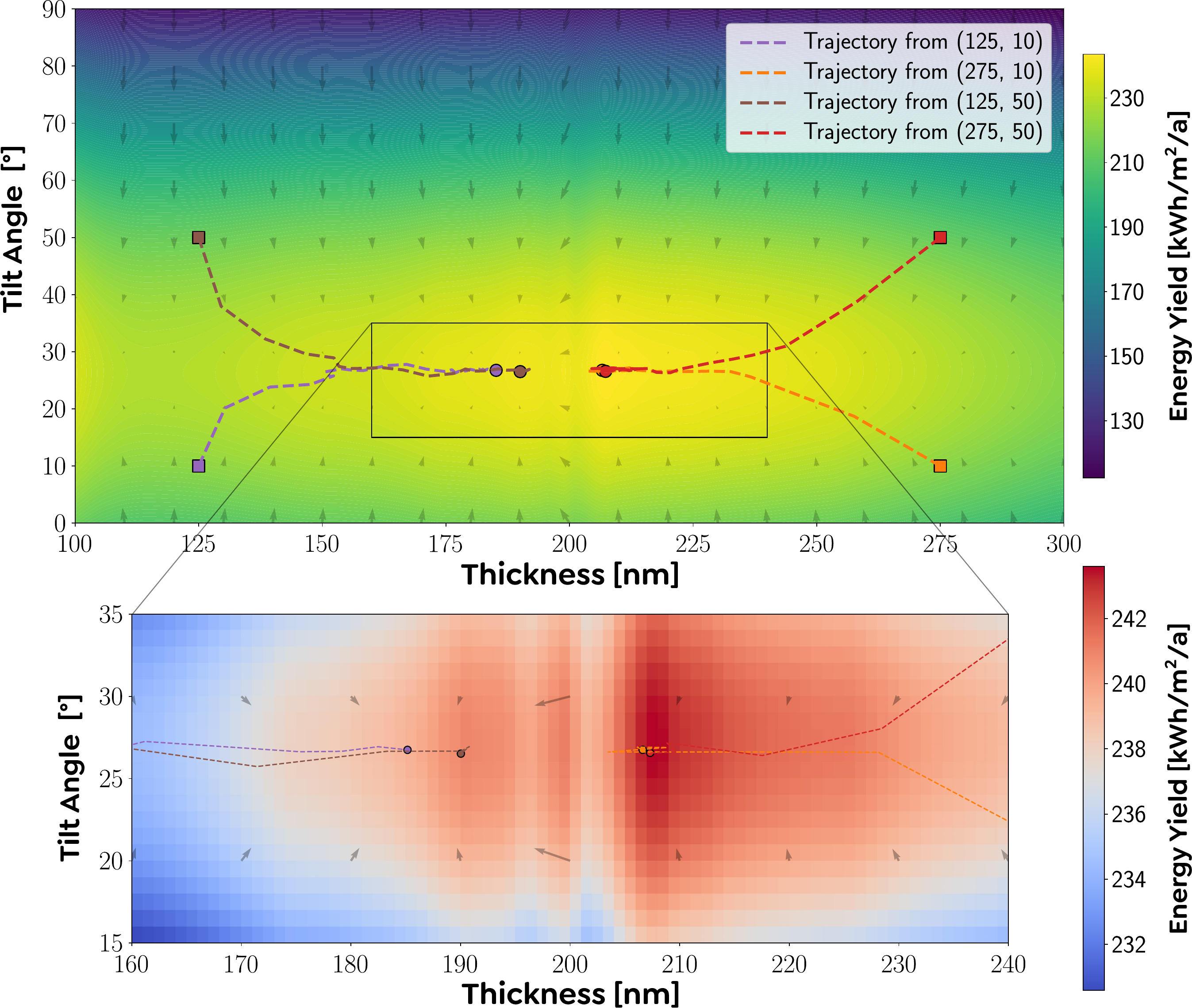}}%
  \caption
    {Gradient-based optimization results for Phoenix over the tilt angle and the photoactive layer thickness. The top panel shows the energy yield (EY) contour plot together with the optimization trajectories starting from four different initial parameter values. Arrows indicate the magnitude and direction of the EY gradient, pointing toward local or global maxima; the optimization paths follow these arrows until converging to a maximum. The bottom panel provides a zoomed-in view of the region surrounding the global maximum to improve visual clarity.}
  \label{fig:optimization}
\end{figure}

\begin{table}[h!]
\centering
\resizebox{\textwidth}{!}{
\begin{tabular}{lccc}
\toprule
\textbf{Location (Climate Type)} 
& \textbf{Optimal Tilt Angle [°]} 
& \textbf{Optimal Thickness [nm]} 
& \textbf{Energy Yield [kWh/m²/a]} \\
\midrule
Honolulu (Tropical)  & 14.8 & 207 & 208.5 \\
Phoenix (Arid)       & 26.6 & 207 & 243.6 \\
Seattle (Temperate)  & 25.8 & 207 & 154.7 \\
\bottomrule
\end{tabular}
}
\caption{Optimal parameters for each location with corresponding energy yield values.}
\label{tab:optimal_params}
\end{table}

The results presented in Table~\ref{tab:optimal_params} highlight key insights into the optimization of solar cell parameters at various locations. While the optimal photoactive layer thickness remains consistent across locations, determined by the intrinsic properties of the material, the tilt angle varies significantly. This makes the reported cell more accessible and scalable for various regions. This variation is driven by the location-dependent solar position, which influences the angle of incidence and seasonal variations in sunlight. The photoactive layer thickness must balance efficient light absorption and charge transport, avoiding the pitfalls of being either too thin or too thick, which could lead to suboptimal EY. 

The development of a digital twin like $\text{Sol}(\text{Di})^2\text{T}$ proves valuable for tuning the parameters of solar cells based on location-specific conditions. By simulating and optimizing parameters like tilt angle and material thickness, the digital twin enables precise adjustments to maximize EY for different geographic regions. This approach ensures a cost-effective and efficient methodology for solar system deployment in diverse environments, aligning with the growing demand for location-specific renewable energy solutions.

\section{Conclusion}
In this work, we develop an innovative and comprehensive simulation framework $\text{Sol}(\text{Di})^2\text{T}$ to calculate the EY of PV devices, exemplified and discussed for an OPV device. Our approach integrates morphological simulations, optical modeling, electrical details, and environmental conditions within a unified framework. Unlike previous approaches, which frequently treat these aspects separately, our digital twin enables more precise EY calculations and captures the impact of nanoscale degrees of freedom on the macroscopic prediction of the EY. 

The developed tool $\text{Sol}(\text{Di})^2\text{T}$ is established with a differentiable structure, enabling gradient-based optimization. With this, we can make a step towards inverse design strategies that optimize solar cell devices under specific boundary conditions. Furthermore, the digital twin establishes a platform linking simulation and experiment. 

We demonstrate $\text{Sol}(\text{Di})^2\text{T}$ for an organic solar cell with PM6:Y6 as the photoactive material. We illustrate how key design parameters and installation layouts are optimized for a given setting. The work indicates a route towards tailored solar cell architectures for specific applications, by fine-tuning controllable degrees of freedom during the fabrication process. Possibly, individual improvements may be small, but they can accumulate to a significant effect in total. 

Beyond personalized PV design, we foresee a promising impact of this digital twin on future applications for emerging PV applications. It enables the combination of EY considerations with additional aspects, such as color appearance for building-integrated PV or crop yield for agro-photovoltaics. This becomes especially relevant in the context of complex, multi-benefit PV applications, where intuition fails to guide optimization. In such cases, modern approaches from the field of ML, as used in our contribution, can help balance competing requirements. They help to find and identify engineering optima within the rapidly growing parameter space.  

We emphasize that $\text{Sol}(\text{Di})^2\text{T}$ is deliberately designed as a minimal, modular platform that integrates expertise from different domains and fosters interdisciplinary collaboration. Released as open-source code, the current implementation of $\text{Sol}(\text{Di})^2\text{T}$ provides a baseline tool that can be extended by individual researchers to suit their specific modeling needs. This opens the perspective for research groups to further develop their tools within individual sub-domains and to integrate them into the unified framework. Crucially, the digital twin enables systematic exploration of how modifications at descriptive levels propagate to influence overall device performance. With broader community adoption, this could become a powerful framework driving theoretical, computational, and device-specific solar cell research, well into the upcoming years. Diversifying the application scenarios beyond organic solar cells, \textit{e.g.}, to perovskite solar cells, will further expand its applicability.   

Overall, our digital twin $\text{Sol}(\text{Di})^2\text{T}$ represents the next step in solar cell design, readily adaptable across a broad range of emerging PV technologies.

\medskip
\textbf{Supporting Information} 
Supporting Information is available from the Wiley Online Library or from the authors.

\textbf{Data availability}
The datasets used in this work are available on the following GitHub repository: \url{https://github.com/aimat-lab/SolDi2T}.

\textbf{Code availability}
The source code used in this work is available on the following GitHub repository: \url{https://github.com/aimat-lab/SolDi2T}.

\medskip
\textbf{Competing interests}
All authors declare no financial or non-financial competing interests. 

\medskip
\textbf{Acknowledgment} \par 
This work has been financially supported by the Helmholtz Association in the framework of the Solar Technology Acceleration Platform “SolarTAP”. 
M.K., P.F., and C.R. acknowledge support by the Deutsche Forschungsgemeinschaft (DFG, German Research Foundation) under Germany’s Excellence Strategy via the Excellence Cluster 3D Matter Made to Order (EXC-2082/1-390761711) and from the Carl Zeiss Foundation via the CZF-Focus@HEiKA Program. M.K. and C.R. acknowledge funding by the Volkswagen Foundation. M.K. and C.R. acknowledge support by the state of Baden--Württemberg through bwHPC and the German Research Foundation (DFG) through grant no. INST 40/575-1 FUGG (JUSTUS 2 cluster). M.L.S. and J.D.F. acknowledge the Karlsruhe School of Optics and Photonics (KSOP). 
Parts of this work were conducted on the HoreKa supercomputer funded by the Ministry of Science, Research, and the Arts Baden--Württemberg and by the Federal Ministry of Education and Research. The Airmass irradiance dataset was obtained from the U.S. Department of Energy (DOE)/NREL/ALLIANCE.
J.H., O.J.J.R., and Y.A. acknowledge support by the Deutsche Forschungsgemeinschaft (DFG, German Research Foundation, Project 449539983), the European Commission (H2020 Program, Project 101008701/EMERGE). 
S.R. acknowledges that this work is funded by the Deutsche Forschungsgemeinschaft (DFG, German Research Foundation) – project number 539945054. 
U.P. and S.O. acknowledge financial support by the Helmholtz Association: Project Zeitenwende (“accelerated transfer of the next generation of solar cells to mass production”), program-oriented funding IV of the Helmholtz Association (Materials and Technologies for the Energy Transition, Topic 1: Photovoltaics and Wind Energy, Code: 38.01.04). C.S. gratefully acknowledges financial support by the Vector Foundation.

\medskip
\textbf{Author contributions}

\textbf{M.L.S.}: Conceptualization (equal), Data curation (equal), Project administration (lead), Formal analysis - Optical Simulations (lead), Investigation - Optical Simulations (lead), Methodology - Optical Simulations (lead), Software - Optical Simulations (equal), Validation (equal), Visualization (lead), Writing – original draft (lead); 
\textbf{H.M.}: Conceptualization (equal), Data curation (equal), Formal analysis - Optimization (lead), Investigation - Optimization (lead), Methodology - Optimization (lead), Project administration (equal), Software - Optimization (lead), Validation (equal), Visualization (equal), Writing – original draft (lead);
\textbf{J.D.F}: Formal analysis - Optical Simulations (equal), Investigation - Optical Simulations (equal), Methodology - Optical Simulations (equal), Software - Optical Simulations (equal), Visualization (equal)
\textbf{B.Z.}: Software - Optical Simulations (supporting);
\textbf{M.K.}: Formal analysis - Optical Simulations (equal), Investigation - Optical Simulations (equal), Methodology - Optical Simulations (equal), Software - Optical Simulations (equal), Visualization (equal), Writing – original draft (equal);
\textbf{U.W.P}: Funding Acquisition (equal), Investigation - Energy Yield Simulation (equal), Supervision (equal), Writing – review \& editing (supporting);
\textbf{S.O.}: Formal analysis - Energy Yield Simulation (lead), Investigation - Energy Yield Simulation (lead), Methodology - Energy Yield Simulation (lead), Software - Energy Yield Simulation (lead), Visualization (equal), Writing – original draft (equal),  Writing – review \& editing (equal);
\textbf{O.J.J.R.}: Investigation - Morphology Simulation (equal), Supervision (equal), Writing – original draft (supporting), Writing – review \& editing (equal); 
\textbf{Y.A.}: Formal analysis - Morphology Simulation (lead), Investigation - Morphology Simulation (lead), Methodology - Morphology Simulation (lead), Software - Morphology Simulation (lead), Visualization (equal), Writing – original draft (equal),  Writing – review \& editing (equal);
\textbf{J.H}: Funding Acquisition (equal), Supervision (supporting), Writing – review \& editing (supporting); 
\textbf{T.K.}: Funding Acquisition (equal), Investigation - Electrical Simulation (equal), Supervision (equal), Writing – review \& editing (equal);
\textbf{S.R}: Formal analysis - Electrical Simulation (equal), Investigation - Electrical Simulation (equal), Methodology - Electrical Simulation (equal), Software - Electrical Simulation (equal), Visualization (supporting), Writing – original draft (equal), Writing – review \& editing (equal);
\textbf{C.D.}: Investigation - Electrical Simulation (equal), Methodology - Electrical Simulation (supporting), Software - Electrical Simulation (supporting), Visualization (supporting), Writing – original draft (equal), Writing – review \& editing (equal);
\textbf{E.K.}: Investigation - Electrical Simulation (supporting), Methodology - Electrical Simulation (supporting), Software - Electrical Simulation (supporting);
\textbf{C.S.}: Funding Acquisition (equal), Conceptualization (equal), Validation (equal), Writing – review \& editing (equal); 
\textbf{M.H.}: Validation (equal), Writing – review \& editing (equal);
\textbf{A.C.}: Funding Acquisition (equal), Conceptualization (equal), Validation (equal), Writing – review \& editing (equal);
\textbf{K.F.}: Conceptualization (equal), Validation (equal), Writing – review \& editing (equal);
\textbf{K.J.}: Investigation - Optical Simulations (supporting), Writing – review \& editing (equal);
\textbf{P.F.}: Funding Acquisition (equal), Investigation - Optimization (equal), Supervision (lead), Writing – original draft (equal), Writing – review \& editing (equal);
\textbf{C.R.}: Funding Acquisition (equal), Investigation - Optical Simulations (equal), Project administration (equal), Supervision (lead), Writing – original draft (equal), Writing – review \& editing (equal)

\medskip

\printbibliography 

\end{document}